\documentclass[]{pasj01}
\Received{\today}
\Accepted{$\langle$acception date$\rangle$}
\Published{$\langle$publication date$\rangle$}
\SetRunningHead{Akahori et al.}{ATCA 16 cm Observations of CIZA J1359-48}

\begin{document}

\title{ATCA 16 cm Observation of CIZA J1358.9-4750: Implication of Merger Stage and Constraint on Non-Thermal Properties}

\author{
Takuya \textsc{Akahori}\altaffilmark{1,2}\thanks{Corresponding author: takuya.akahori@nao.ac.jp}, Yuichi \textsc{Kato}$^{3}$, Kazuhiro \textsc{Nakazawa}$^{3,4,5}$, Takeaki \textsc{Ozawa}$^{1,2}$, Liyi \textsc{Gu}$^6$, Motokazu \textsc{Takizawa}$^7$, Yutaka \textsc{Fujita}$^8$, Hiroyuki \textsc{Nakanishi}$^1$, Nobuhiro \textsc{Okabe}$^{9,10}$, and Kazuo \textsc{Makishima}$^{3,11}$
}

\altaffiltext{}{
$^1$Graduate School of Science and Engineering, Kagoshima University, 
1-21-35, Korimoto, Kagoshima, Kagoshima 890-0065, Japan \\
$^2$National Astronomical Observatory Japan, 2-21-1 Osawa, Mitaka, Tokyo 181-8588, Japan \\
$^3$Department of Physics The University of Tokyo, 7-3-1 Hongo, Bunkyo-ku, Tokyo, 113-0033, Japan \\
$^4$Center for Experimental Studies, Kobayashi-Maskawa Institute for the Origin of Particles and the Universe, Furo-cho, Chikusa-ku, Nagoya, Aichi 464-8602, Japan \\
$^5$Division of Particle and Astrophysical Science, Graduate School of Science, Nagoya University, Furo-cho, Chikusa-ku, Nagoya, Aichi 464-8602, Japan \\
$^6$SRON Netherlands Institute for Space Research, Sorbonnelaan 2, 3584 CA Utrecht, the Netherlands \\
$^7$Department of Physics, Yamagata University, Kojirakawa-machi 1-4-12, Yamagata 990-8560, Japan \\
$^8$Department of Earth and Space Science, Graduate School of Science, Osaka University, Toyonaka, Osaka 560-0043, Japan \\
$^9$Department of Physical Science, Hiroshima University, 1-3-1, Kagamiyama, Higashi-Hiroshima, Hiroshima 739-8526, Japan \\
$^{10}$Hiroshima Astrophysical Science Center, Hiroshima University, Higashi-Hiroshima, Kagamiyama 1-3-1, 739-8526, Japan \\
$^{11}$MAXI Team, RIKEN, 2-1 Hirosawa, Wako, Saitama 351-0198, Japan \\
}

%%\email{takuya.akahori@nao.ac.jp}


\maketitle

%%%%%%%%%%%%%%%%%%%%%%%%%%%%%%%%%%%%%%%%%%%
%%%%%%%%%%%%%%%%%%%%%%%%%%%%%%%%%%%%%%%%%%%
\begin{abstract}

We report the Australia Telescope Compact Array 16 cm observation of CIZA J1358.9-4750. Recent X-ray studies imply that this galaxy cluster is composed of merging, binary clusters. Using the EW367 configuration, we found no significant diffuse radio emission in and around the cluster. An upper limit of the total radio power at 1.4 GHz is $\sim 1.1 \times 10^{22}$ Watt/Hz in $30$ square arcminutes which is a typical size of radio relics. It is known that an empirical relation holds between the total radio power and X-ray luminosity of the host cluster. The upper limit is about one order of magnitude lower than the power expected from the relation. Very young ($\sim 70$~Myr) shocks with low Mach numbers ($\sim 1.3$), which are often seen at an early stage of merger simulations, are suggested by the previous X-ray observation. The shocks may generate cosmic-ray electrons with a steep energy spectrum, which is consistent with non-detection of bright ($>10^{23}$~Watt/Hz) relic in this 16 cm band observation. Based on the assumption of energy equipartition, the upper limit gives the magnetic-field strength below $0.68f(D_{\rm los}/{\rm 1~Mpc})^{-1}(\gamma_{\rm min}/200)^{-1}$~$\mu$G, where $f$ is the cosmic-ray total energy density over the cosmic-ray electron energy density, $D_{\rm los}$ is the depth of the shock wave along the sightline and $\gamma_{\rm min}$ is the lower cutoff Lorentz factor of the cosmic-ray electron energy spectrum.

\end{abstract}

%%%%%%%%%%%%%%%%%%%%%%%%%%%%%%%%%%%%%%%%%%%
%%%%%%%%%%%%%%%%%%%%%%%%%%%%%%%%%%%%%%%%%%%
\section{Introduction}
\label{section1}

In the large-scale structure formation, merging galaxy clusters release their huge gravitational energy into thermal energy of the intracluster medium (ICM) (e.g. \cite{sar02, mar07}). A merger sequence of two clusters can be divided into (i) the early stage where the two clusters are getting close (e.g., Abell 399/401, \cite{fuj08}), and (ii) the late stage where they are receding from each other (e.g., Abell 3667, \cite{nak09}); an intermediate stage between the above two phases occurs in a short time-scale because of high-speed crossing of the two gravitational potential cores. Hydrodynamic simulations suggest that shock waves with position-dependent Mach numbers, $M$, arise in the merger (e.g., \cite{ryu03, tak08, ay10}). The shocks with $M<2$ appear in the linking region of the two clusters in the early stage, while the shocks with larger Mach numbers, $2<M<4$, appear ahead of the two receding cores in the late stage because the shocks propagate to cooler gas at cluster outskirts.

The merger stage is, therefore, one of the essential information to understand the nature of cluster merger. However, although shock waves and their Mach numbers are important clues for elucidating the merger stage, they are studied only in a limited number of bright X-ray clusters (e.g., \cite{fi10, ak13, ita15}). Even though we know the presence of shocks, the merger stage is controversial in some clusters (e.g., RX J1347.5-1145, \cite{kit16}). Significant effort has been ongoing over the last years to determine the merger stage by morphology of X-ray surface brightness, analysis of weak gravitational lensing, optical spectroscopy, and galaxy distribution (e.g., \cite{oka08, oka11, daw13, daw15, oka15, jee16}). For example, separation of a cD galaxy from the peak of X-ray surface brightness is frequently seen in the late stage of merger (Abell 2163, \cite{oka11}).

Radio observation provides us with complementary information of the shock wave and the merger stage (e.g. \cite{oza15}). Many clusters possess diffuse radio emissions, which are in general classified into radio halos, mini-halos, and relics, according to their size, location, morphology, and so on (see e.g. \cite{fer12} for a recent review). The connection between diffuse radio emission and dynamical state of the system has been discussed (e.g. powerful radio halos in underluminous X-ray clusters; \cite{gio11}, and giant radio halos in cool core systems; \cite{bon14}). Double-arc shapes seen in several radio relics suggest the presence of shock waves which appear in the late stage of merger, although there are also radio relics having more roundish shapes. The origin of such diversity on the relics has been intensively studied in the literature (see \cite{fer12}). 

The cluster diffuse radio emission is likely synchrotron radiation, suggesting the presence of intracluster magnetic field and cosmic-ray electrons (CRes). Because GeV CRe emitting synchrotron radiation in GHz band has a short lifetime, the emission would be associated with fresh CRes which are recently accelerated (e.g., \cite{sar86, en98, ct02}). For the merger shocks, diffusive shock acceleration (DSA) has been studied in the literature (e.g., \cite{be87, kr13, vb14}). This theory can link the spectral index of radio emission with the shock Mach number. However, the standard DSA (the first-order Fermi acceleration) suffers from a low efficiency of acceleration, particularly at a shock with a low Mach number. To overcome this injection problem, another mechanisms have been studied such as shock drift acceleration (SDA, e.g. \cite{ma11, cs14, guo14}; \cite{ma15}) and second-order Fermi (re-)acceleration \citep{fuj15, fuj16, kan17}. \citet{van17} found a radio emission and interpreted it as synchrotron radiation from the CRes which were originally injected by an AGN and were recently re-accelerated by a cluster merger shock.

Various mechanisms take place in the structure formation; shock heating, eddy cascading (turbulence), cosmic-ray acceleration, and magnetic-field amplification, as well as radiative cooling, AGN feedback, and metal enrichment. Therefore, observational diagnostics of energy budgets in galaxy clusters is crucially important to understand the structure formation history. Thermal and non-thermal energies in the ICM can be partly estimated from X-ray continuum emission and synchrotron emission, respectively. It is known that a correlation holds between the radio power, $P$, and the X-ray luminosity, $L_{\rm X}$, of the host cluster. This $L_{\rm X}$--$P$ relation applies to both radio halos and relics (e.g., \cite{bru07, bru09, rud09, cas10, ens11, fer12, xu12, gov13, cas13, deg14, kal15}) and even to radio mini-halos in cool-core clusters \citep{git15}.

Although many late-stage clusters have been observed, there are only several candidates of early-stage clusters, and little is known about shock waves, radio emission, the radio spectral index, and the $L_{\rm X}$--$P$ relation in the early stage (e.g. \cite{ak16}). In the present paper, we report the first 16 cm (1--3 GHz) observations of CIZA J1358.9-4750 (hereafter CZ1359), which is a candidate of an early-stage cluster. We discuss the nature of CZ1359 such as merger scenario and particle acceleration. This paper is organized as follows. We describe the observations and data reduction in Sections 2 and 3, respectively. Results are shown in Section 4. Our discussion is presented in Sections 5, followed by concluding remarks in Section 6.

%%%%%%%%%%%%%%%%%%%%%%%%%%%%%%%%%%%%%%%%%%%
%%%%%%%%%%%%%%%%%%%%%%%%%%%%%%%%%%%%%%%%%%%
\section{Radio Observations}
\label{section2}

%%%%%%%%%%%%%%%%%%%%%%%%%%%%%%%%%%%%%%%%%%%
\subsection{The Target}
\label{section2.1}

CZ1359 is a nearby (redshift $z\sim 0.07$) galaxy cluster listed in the X-ray cluster catalog, CIZA (the Clusters in the Zone of Avoidance, \cite{ebe02, koc07}). This object exhibits two X-ray surface brightness enhancements at South-East and North-West directions with a separation of $14' \sim 1.2$~Mpc. The enhancements are likely sub-clusters and the separation is $\sim 0.7$--$0.8$ times smaller than the virial radius ($r_{200}$, \cite{kat15}) of each sub-cluster. Each X-ray core possesses a giant elliptical galaxy, which does not significantly deviate from the X-ray center.

\citet{kat15} observed CZ1359 with the Suzaku X-ray telescope, and found that the ICM in the linking region between the cores is $\sim 30$~\% hotter than the ICM around the cores. Moreover, they discovered a temperature jump in the region and confirmed an associated X-ray brightness jump in an archival XMM-Newton image. Supposing the Rankine-Hugoniot condition to the largest temperature increase, they estimated the Mach number of the shock, $M=1.32 \pm 0.22$. This number is broadly consistent with a prediction by numerical simulations ($M\sim 1.5$, \cite{ay08}; 2010). 

The coincidental position of giant elliptical galaxies with X-ray cores, the X-ray brightness enhancement, and the X-ray temperature jumps all suggest that CZ1359 is in the early stage of merger, though further evidence such as radio emission is helpful to understand its thermal and dynamical state. There are no scientific centimeter observation covering the CZ1359 field, except the 843~MHz SUMSS shallow survey \citep{bock99}. There are also two 150 MHz surveys to be publicly available: the TGSS \citep{int17} and the GLEAM \citep{hur17}. All of these surveys detected several compact sources in the CZ1359 field (Section \ref{section4}), but no diffuse radio emission (Lijo Thomas George and Ruta Kale, private communication).

%%%%%%%%%%%%%%%%%%%%%%%%%%%%%%%%%%%%%%%%%%%
\subsection{The Observation}
\label{section2.2}

\begin{table*}
\tbl{The observation log of CZ1359 with the ATCA.}{%
\begin{tabular}{llllllll}
\hline
\hline
pointing & right ascension & declination & date & period (UT) & frequency (MHz) & bandwidth (MHz) & exposure (min)\\
\hline
NE & 13:59:10.0 & -47:41:00.0 & June 06, 2014 & 04:58 -- 16:37 & 1100-3100 & 2048 & 560\\
NW & 13:58:10.0 & -47:40:00.0 & June 08, 2014 & 05:02 -- 17:06 & 1100-3100 & 2048 & 595\\
SE & 13:59:10.0 & -47:52:00.0 & June 09, 2014 & 04:30 -- 16:36 & 1100-3100 & 2048 & 595\\
SW & 13:58:00.0 & -47:51:00.0 & June 10, 2014 & 04:31 -- 16:46 & 1100-3100 & 2048 & 560\\
\hline
\end{tabular}}\label{t01}
\end{table*}

In order to explore possible radio halos, mini-halos, and relics in the CZ1359 field, we carried out a deep centimeter observation of the entire CZ1359 field which extends $\sim 30'$ in diameter. The observation was made with the Australia Telescope Compact Array (ATCA). We adopted the 16 cm band (from 1.1~GHz to 3.1~GHz) which gives the field-of-view of $22'.3$ at the center frequency, 2.1~GHz. The observation required a trade-off between (i) an array configuration as compact as possible to gain a better sensitivity for diffuse emission, and (ii) a beam size of $\sim 1'$ or better so as to identify arc-shape structures of potential relics. We adopted the EW367 array configuration in which all six antennas were mounted in the east-west track. The minimum baseline was 46~m and the largest well-imaged structure at 2.1 GHz is $\sim 20'$ \footnote{The largest well-imaged structure of 1-day observation is $\sim 230''$ at 6 cm with the 750 m configuration (section 1.7.1 of ATCA Users Guide, http://www.narrabri.atnf.csiro.au/observing/users\_guide/html/atug.html). It scales $230''*2*16/6\sim 20'$ at 16 cm with the 367~m configuration.}. The maximum baseline was 367~m without the distant antenna 6 (CA06), and 4408~m with CA06. Resultant beam position angles and beam major/minor axes at 2.1 GHz were $-89$ degree and $\sim 23'' \times 20''$ with CA06, and $-1.6$ degree and $\sim 137'' \times 91''$ without CA06, respectively.

We adopted four pointings to cover the CZ1359 field. About a half of each primary beam overlaps the other beams. The observations were performed in June 6 (north-east, NE), June 8 (north-west, NW), June 9 (south-east, SE), and June 10 (south-west, SW) all in 2014 (PI: T. Akahori, ID: C2916). The parameters of these pointings are listed in Table \ref{t01}. Observation time for each pointing was 12 hours in total including the overhead of standard calibration and antenna slew time. On-target exposure time were 9.33, 9.92, 9.92, and 9.33 hours in total for NE, NW, SE, and SW, respectively. Such a long tracking provided us with good hour angle coverage\footnote{The ATCA uses dual linear orthogonal feeds to measure two orthogonal linear polarizations simultaneously. The position angle of the polarization splitter is stationary with respect to the altra-azimutual-mounted antennas and so rotates on the sky.}. The system temperature was between 38 K (meridian passage) and 50 K (rising and setting) and those were stable during the observing days. 

In each day, we observed the calibrator 0823-500 for 10 minutes at the start of the schedule, the band-pass calibrator 1934-638 (12.6 Jy at 2.1 GHz, a compact source, no linear polarization, where 1~Jy = $10^{-26}$~${\rm Watt/Hz/m^2}$) for 10 minutes at the middle and close of the schedule, and the gain/phase calibrator 1421-490 (7.38 Jy at 2.1 GHz, the distance from the target is 4.48 degree) for 3 minutes every 35 minutes of on-target observations. The 1M-0.5k Compact Array Broadband Backend (CABB, \cite{wil11}) receiver mode was selected to allow full spectro-polarimetry observations. The raw spectral channel is 1 MHz and the bandwidth is 2048 MHz.

%%%%%%%%%%%%%%%%%%%%%%%%%%%%%%%%%%%%%%%%%%%
%%%%%%%%%%%%%%%%%%%%%%%%%%%%%%%%%%%%%%%%%%%
\section{Data Reduction}
\label{section3}

We performed the standard ATCA data reduction by using the MIRIAD software (version 1.5) as follows. All data were loaded by the MIRIAD task \textsc{atlod} with options \textsc{birbie}, \textsc{rfiflag}, \textsc{xycorr}, and \textsc{noauto}. Each 40 channels at the band edges that are likely affected by the bandpass rolloff were removed by \textsc{uvflag}. Data were then split into subsets for the band-pass calibrator, the gain/phase calibrator, and the target, using \textsc{uvsplit}. For each subset, we performed \textsc{uvflag}, \textsc{pgflag}, and \textsc{blflag} to flag out radio frequency interferences (RFIs). These were done iteratively until RFI signals disappear in visibility plots. The band-pass solution was made by \textsc{mfcal} using the subset of the band-pass calibrator. The solution was transferred to the gain/phase calibration by \textsc{gpcopy}, and then \textsc{gpcal} was carried out to obtain the gain and phase solutions. The error of the absolute value was computed by bootstrap estimation using \textsc{gpboot}. Finally, all solutions were applied to the target data using \textsc{gpcopy} and \textsc{gpaver}.

The visibility data were transformed into images using \textsc{invert}, where the robustness parameters, $r=0.5$ and $2.0$, were chosen to optimize the images for the studies of compact and diffuse sources, respectively. We created the images including CA06 which gives the longest baselines and improves the angular resolution. We also created the images excluding CA06, because it makes the image rms noise level worse due to the large side-lobes caused by large uv gaps between the longest baselines and the others. The deconvolution was carried out using \textsc{mfclean}, where we employed the best iteration number which minimize the rms noise on the cleaned image. Finally, four pointing images were linearly combined into a mosaic image using \textsc{linmos}.

\begin{figure}[tp]
\begin{center}
\FigureFile(70mm,70mm){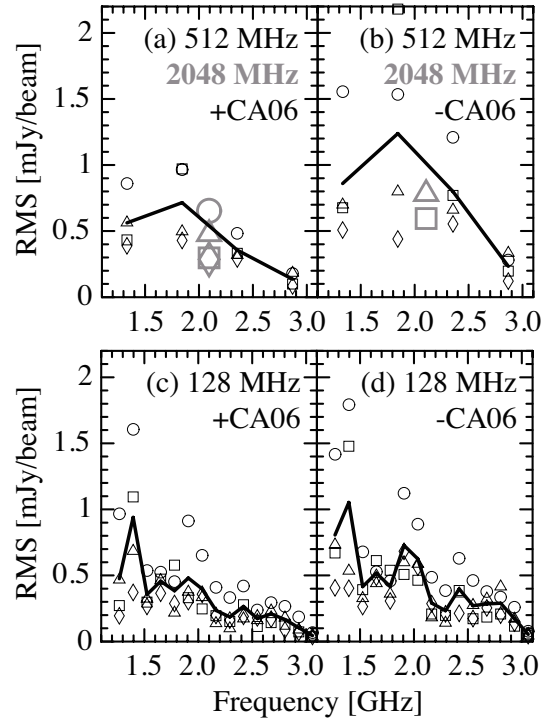}
\end{center}
\caption{
(a) Image rms noise levels for the NE (circles), NW (boxes), SE (triangles), and SW (diamonds) pointings with 512 MHz (black) and 2048 MHz (gray) bandwidth. The solid line is the average for the four pointings of 512 MHz bandwidth. (b) The same as (a) but without CA06 baselines. Note that rms noise levels of NE (51.26 mJy/beam) and SW (4.730 mJy/beam) for a 2048 MHz bandwidth are outside the frame of (b). (c)(d) The same as (a) and (b) respectively, but with a 128 MHz bandwidth. 
}
\label{f01}
\end{figure}

The Stokes I images were successfully reduced and the flux scale was verified using a SUMSS source (Appendix A). The 128, 512, and 2048 MHz bandwidth images were made to compare the rms noise levels. Here, in the 128 MHz case, we did not consider the bottom (1140 MHz) and top (3060 MHz) bands, since the noise levels were high due to RFIs. Figure~\ref{f01} summarizes the rms noise levels we have achieved with the robustness $r=2.0$. Typical noise levels were 0.5 mJy/beam (1.5--2 GHz) and 0.3 mJy/beam (2--3 GHz). Meanwhile, Stokes Q and U images have no significant signal. We did not obtain reliable noise levels of Stokes Q and U because of artefacts in the images.

The typical noise levels were several times worse than the theoretical rms noise level (e.g., 0.038 mJy/beam at 2036 MHz with a 128 MHz bandwidth). This is thought to be mostly due to a very bright source PMN J1401-4733 (or SUMSS J140158-473336) located at the edge of each pointing image. The total intensity of this source, 0.433 Jy/beam at 1524 MHz, was $\sim 1000$ times higher than the image rms noise level, and this strong signal inevitably produces frequency-dependent sidelobes. Actually, the sidelobe effect makes the noise highest in the NE images, and increasing the bandwidth above 128 MHz did not improve the image rms noise level, as shown in Fig.~\ref{f01}. The sidelobe effect cannot be removed by tuning the robustness and taper parameters in imaging, nor by iterating CLEAN algorithm. We also performed the self-calibration, but the noise level remains within $\pm~10-20$~\%, depending on the direction (NE, NW, SE, SW) and the frequency. Therefore, hereafter we only show our results without taking the self-calibration.

%%%%%%%%%%%%%%%%%%%%%%%%%%%%%%%%%%%%%%%%%%%
%%%%%%%%%%%%%%%%%%%%%%%%%%%%%%%%%%%%%%%%%%%
\section{Result}
\label{section4}

We first report compact sources found in the CZ1359 field, and then explore diffuse sources. As mentioned, visibility data were imaged with the robustness $r=0.5$ for compact sources and $r=2.0$ for diffuse sources to optimize each detection.

%%%%%%%%%%%%%%%%%%%%%%%%%%%%%%%%%%%%%%%%%%%%
\subsection{Compact Sources}

\begin{table*}
\tbl{Summary of compact sources in the CZ1359 field.}{%
\begin{tabular}{lrrrrrrrr}
\hline
\hline
ID & right ascension & declination & this work$^\dagger$ & $\alpha_{\rm obs}$ & TGSS$^\ddagger$ & SUMSS$^*$ & 2MASX & $\alpha_{\rm obs}^{\rm SUMSS+ATCA}$ \\
& (error in arcsec) & (error in arcsec) & mJy/beam & & mJy/beam & mJy/beam & &\\
\hline
A & 13:59:46.536 (1.139) & -47:54:14.37 (0.975) &  6.96 $\pm$ 0.44 & 0.33 & 18.32 & 13.2 $\pm$ 1.2 & --- & 0.67\\
B & 13:59:37.828 (1.342) & -47:53:52.25 (1.142) &  6.45 $\pm$ 0.47 & 0.94 & 16.16 & --- & --- & \\
C & 13:59:31.085 (1.455) & -47:54:06.73 (1.239) &  7.16 $\pm$ 0.92 & 0.98 & 21.76 & --- & Yes & \\
D & 13:59:24.841 (1.613) & -47:50:19.07 (1.431) &  3.39 $\pm$ 0.26 & 1.86 & --- & --- & Yes & \\
E & 13:59:22.770 (1.625) & -47:49:05.25 (1.411) & 6.53 $\pm$ 0.57 & 0.95 & ---& 11.2 $\pm$ 1.1 & Yes & 0.57\\
F & 13:59:19.316 (1.559) & -47:36:41.85 (1.344) &  4.68 $\pm$ 0.51 & 0.22 & 34.83 & 15.2 $\pm$ 1.1 & --- & 1.24\\
G & 13:59:03.488 (1.987) & -47:51:33.62 (1.709) &  2.99 $\pm$ 0.26 & 0.66 & --- & --- & Yes$^{\S}$ & \\
H & 13:58:53.790 (1.151) & -47:30:55.74 (0.985) &  6.89 $\pm$ 0.26 & 0.63 & 17.39 & 13.7 $\pm$ 1 & --- & 0.72\\
I & 13:57:53.588 (1.382) & -47:37:48.99 (1.186) &  7.49 $\pm$ 0.57 & 0.51 & 45.39 & --- & Yes & \\
J & 13:57:47.991 (1.508) & -47:37:19.39 (1.290) &  8.28 $\pm$ 0.89 & 0.57 & 44.29 & 26.4 $\pm$ 1.3 & --- & 1.22\\
K & 13:57:41.723 (0.975) & -47:53:03.78 (0.827) &  2.94 $\pm$ 0.41 & 0.44 & 16.04 & --- & --- & \\
L & 13:57:33.947 (1.271) & -47:33:33.36 (1.086) &  7.49 $\pm$ 0.34 & 0.52 & 64.88 & 19.4 $\pm$ 1.1 & --- & 1.00\\
\hline
\end{tabular}}\label{t02}
\begin{tabnote}
$^\dagger$Flux at 2100 MHz. $^\ddagger$Flux at 150 MHz. $^*$Flux at 843 MHz. $^{\S}$See also the text.
\end{tabnote}
\end{table*}

\begin{figure}[tp]
\begin{center}
\FigureFile(80mm,80mm){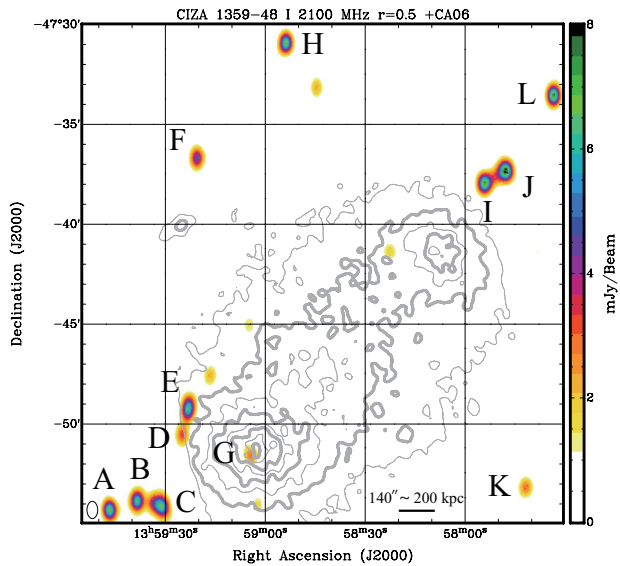}
\end{center}
\caption{
Total intensity map of CZ1359 at 2100 MHz with a 2048 MHz bandwidth and with CA06 baselines. Gray contours show the X-ray surface brightness \citep{kat15}.
}
\label{f02}
\end{figure}

\begin{figure}[tp]
\begin{center}
\FigureFile(80mm,80mm){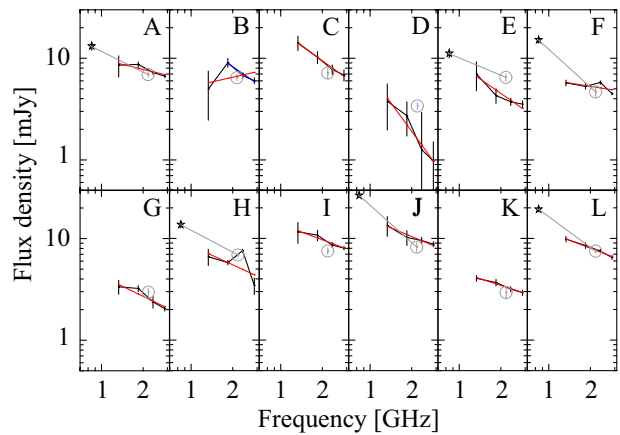}
\end{center}
\caption{
Radio intensity spectra for twelve compact sources in the CZ1359 field. Solid lines and gray circles show the ATCA 512 MHz and 2048 MHz bandwidth data, respectively. Stars indicates the data of SUMSS. Red lines are best-fits of the ATCA 512 MHz data only. The blue line for Source B is the best-fit without the bottom band data (see the text) Gray lines are best-fits of the SUMSS + ATCA 2048 MHz bandwidth data.
}
\label{f03}
\end{figure}

Radio compact sources which have the flux larger than the detection limit ($>5\sigma$) were explored, and twelve reliable compact sources were found. Here, by eye, we rejected several spurious sources (not listed) which were unreliable in the two senses that (i) they appear only in less than half of 128 MHz band images, or (ii) their positions depend on frequencies and move beyond angular resolutions. Locations of the detected compact sources are shown in Fig.~\ref{f02} with alphabetic labels from East to West. Spectra of these sources are shown in Fig.~\ref{f03}.

Table~\ref{t02} summarizes the source coordinate and source peak flux, which were derived by using the MIRIAD task \textsc{imfit} with an option \textsc{object=point}. Also listed is the spectral index, $\alpha_{\rm obs}$, which was obtained from the least square fit of our 512 MHz bandwidth spectra with a simple power-law form
\begin{equation}
I_\nu \propto \nu^{-\alpha_{\rm obs}}, 
\end{equation}
where $\nu$ is the frequency. Only Source B indicates a positive slope (a negative spectral index) with $\alpha_{\rm obs}=-0.31$, though the fit is affected by the bottom-band data which have large errors. Without the bottom band, we obtain $\alpha_{\rm obs}=0.94$ (the blue line in Fig.~\ref{f03}), which is adopted in Table~\ref{t02}.

We checked TGSS images and the SUMSS catalog to see possible counterpart emission, where the search range was set to be 30 arcsec (about one and half beam size) in radius. We also searched for possible association of extragalactic objects using the NASA Extragalactic Database (NED). The results found in TGSS (150~MHz) images, the SUMSS (843~MHz) catalog, and 2MASX (near-infrared) database are listed in Table~\ref{t02}. The spectral index, $\alpha_{\rm obs}^{\rm SUMSS+ATCA}$ was derived using the SUMSS data and our 2048 MHz bandwidth data.

\begin{figure}[tp]
\begin{center}
\FigureFile(80mm,80mm){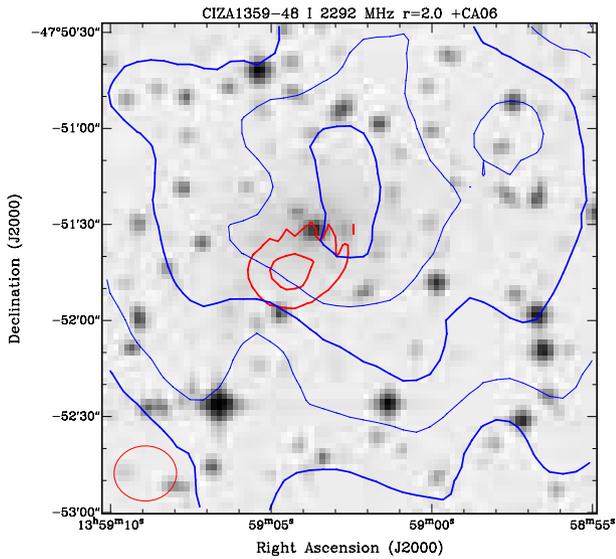}
\end{center}
\caption{
The superposed image near the center of South-East sub-cluster of CZ1359. Background is the DSS optical image. Blue contours are the Suzaku X-ray surface brightness. Red contours are 2.3 GHz radio continuum images for a 128 MHz bandwidth with levels of 1 (5.4 $\sigma$) and 2 (10.7 $\sigma$) mJy/beam. Also shown at the left-bottom is the beam size of the radio image.
}
\label{f04}
\end{figure}

It is found that Source G, which has no TGSS and SUMSS counterpart, is located near the center of the SW cluster, 2XMM J135904.3-475125 (13:59:04.3, -47:51:25). A superposed map at 2292 MHz is shown in Fig.~\ref{f04}, where the coordinate errors of right ascension and declination are $1''.42$ and $1''.23$, respectively, and the beam major/minor axes are $19''.79 \times 17''.23$. The peak position of the radio emission deviates from that of the X-ray by about one beam size ($\sim 20''$ or $\sim 30$~kpc at $z=0.07$). It also deviates from the central galaxy 2MASX J13590381-4751311 (13:59:03.8, -47:51:31) and a nearby galaxy 2MASX J13590223-4751502 (13:59:02.2, -47:51:50) each by about a half beam size. The size of Source G is at least smaller than $\sim 40$~kpc and the spectral index is 0.66. These features could be explained by AGN emission in the cluster or background, whose host galaxy is one of the two 2MASX sources or is too faint to list in the 2MASX catalog. With the spectral index of 0.66, the expected flux density at 150 MHz and 843 MHz are 17.1 mJy and 5.46 mJy, both of which are below the detection limits of TGSS and SUMSS, respectively.

%%%%%%%%%%%%%%%%%%%%%%%%%%%%%%%%%%%%%%%%%%%%
\subsection{Diffuse Sources}

\begin{figure}[tp]
\begin{center}
\FigureFile(80mm,80mm){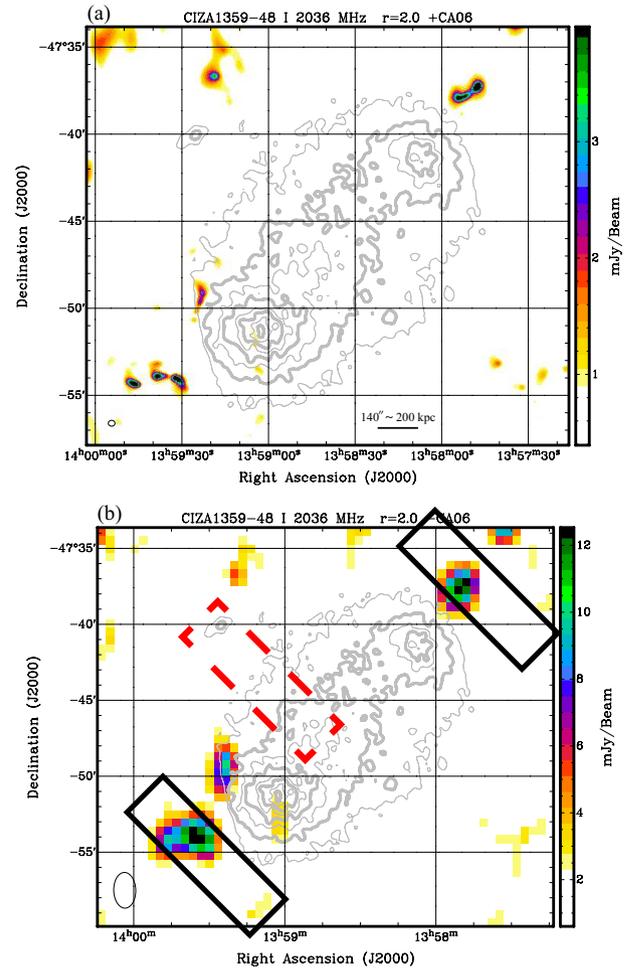}
\end{center}
\caption{
Total intensity map of CZ1359 at 2036 MHz with a 128 MHz bandwidth (a) with CA06 baselines and (b) without CA06 baselines. The color range is shown from 1 $\sigma$ to 10 $\sigma$ rms noise level. Gray contours show the X-ray surface brightness \citep{kat15}. The black-solid and red-dashed boxes indicate $10'\times 3'$ areas in which radio relics are expected in the late- and early-stage of cluster merger, respectively; the latter is associated with the observed X-ray shock front (\cite{kat15}; section 5.4). 
}
\label{f05}
\end{figure}

Fig.~\ref{f05}(a) shows the CA06-included Stokes I image at 2036 MHz with a 128 MHz bandwidth, where gray contours represent the X-ray surface brightness \citep{kat15} for reference. We found only compact sources, and did not detect diffuse sources in the CZ1359 field. The CA06-excluded image is also shown in Fig.~\ref{f05}(b). There are apparently extended emissions, but all of them are consistent with compact sources convolved with a low-resolution beam, or noise. Because the largest well-imaged structure ($\sim 20'$) should be enough to detect possible diffuse emission in the CZ1359 field, it is unlikely that we miss the  diffuse emission due to a lack of short baselines of the interferometry. We have checked all 128, 512, and 2048 MHz bandwidth images, and found no diffuse emission among any of them. 

Let us consider a typical, bright diffuse emission, i.e. radio halo and relic, possessing a surface brightness of $1~\mu$Jy/arcsec$^2$ at 1.4~GHz with $\alpha_{\rm obs} \sim 1$ \citep{fer12}. For the SUMSS at 843 MHz with the beam size of 2295~arcsec$^2$, the brightness becomes 3.8~mJy/beam or only $\sim 3$ times of its sensitivity limit. Meanwhile, for our observation at 2.1 GHz with the beam size of 14126~arcsec$^2$ (without CA06), the brightness becomes 9.5~mJy/beam or $\sim 20$--$30$ times of the achieved sensitivity limit. Therefore, our non-detection firmly excludes the presence of bright diffuse emission in the CZ1359 field. Even a faint case of $0.1~\mu$Jy/arcsec$^2$ is excluded at a $\sim 3\sigma$ level.

\begin{figure}[tp]
\begin{center}
\FigureFile(80mm,80mm){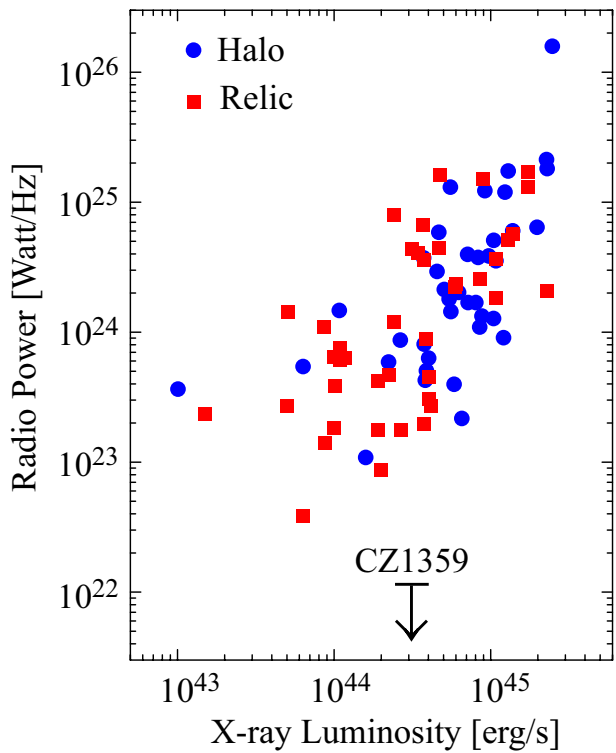}
\end{center}
\caption{
The correlation between the radio power at 1.4 GHz and the 0.1 -- 2.4 keV X-ray luminosity of 39 radio relics (red squares) and 41 radio halos (blue circles), all taken from \citet{fer12}. The black arrow indicates the upper limit of this work for a radio relic derived in Section 4.
}
\label{f06}
\end{figure}

We derive the upper limit of radio power $({\rm Watt/Hz})$ from the integration of the intensity,
\begin{equation}
P = 4\pi D_{\rm L}^2 \int I d\Omega 
\sim 4\pi D_{\rm L}^2 I_{\rm rms}A_{\rm beam}\sqrt{A_{\rm lim}/A_{\rm beam}},
\end{equation}
where $D_{\rm L}$ is the luminosity distance in meter, $I_{\rm rms}$ the noise level in Jy/arcsec$^2$, $A_{\rm beam}$ the beam area in arcsec$^2$, and $A_{\rm lim}$ the upper-limit area in arcsec$^2$. Because the upper-limit area considered below is larger than the beam area, the flux upper-limit is given according to the error propagation; the error on a flux measurement scales with the square root of the ratio of the upper-limit area and beam area. 

We adopt $D_{\rm L}(z=0.07)\sim 9.74 \times 10^{24}$~m for the cosmological parameters $H_0=70$~km~s$^{-1}$~Mpc$^{-1}$, $\Omega_{\rm M}=0.3$, and $\Omega_{\rm \Lambda}=0.7$. The noise level is $I_{\rm rms} = 0.5$~mJy/beam at 1.4~GHz, where the beam size ($A_{\rm beam}$) at 1396 MHz is $955.9$~arcsec$^2$ with CA06 and $29741$~arcsec$^2$ without CA06. The upper-limit area is determined such that radio relics typically extend over a size several times larger than the cores of merging galaxy clusters \citep{fer12}. If we consider a relatively large (i.e. conservative) upper-limit area of $30$ arcmin$^2$ shown in Fig.5, we obtain the upper limit of radio power at 1.4 GHz, $P_{\rm 1.4}=6.34\times 10^{22}$ Watt/Hz with CA06 and $P_{\rm 1.4}=1.14\times 10^{22}$ Watt/Hz without CA06. If we consider a typical, large radio halo with an area of 1 Mpc$^2$ \citep{fer12} or 144 arcmin$^2$ in the CZ1359 field, we obtain $P_{\rm 1.4}=1.39\times 10^{23}$ Watt/Hz with CA06 and $2.49 \times 10^{22}$ Watt/Hz without CA06.

Figure~\ref{f06} shows the $L_{\rm X}$--$P_{\rm 1.4}$ relation taken from \citet{fer12}. The arrow depicts the upper limit of CZ1359, $P_{\rm 1.4}=1.14\times 10^{22}$ Watt/Hz, against the X-ray luminosity, $L_{\rm X}=3.14\times 10^{44}$ erg/s, derived from the Suzaku X-ray observation \citep{kat15}. Here, so as to compare with the other ROSAT data in \citet{fer12}, we derived the 0.1-2.4 keV luminosity from the bolometric luminosity, temperature, and metal abundance. We summed up the X-ray luminosities of the two sub-clusters as the total X-ray luminosity of the system, following studies of other clusters. 

The upper limit of CZ1359 falls by one to two orders of magnitude lower than the $L_{\rm X}$--$P_{\rm 1.4}$ relation. This observation first determined a strong upper limit of radio emission for a known intracluster shock with $M\sim 1.3$. To date, there are several observations which introduce similar upper limits of Mpc-scale radio emission with the radio power of $\lesssim 10^{23}$ Watt/Hz around the X-ray luminosity of $\sim 10^{44}$--$10^{45}$ erg/s (not shown since it makes the plot crowded; see e.g. \cite{kal15}). The upper limit of CZ1359 is a factor of several lower than the limits obtained in previous works.

%%%%%%%%%%%%%%%%%%%%%%%%%%%%%%%%%%%%%%%%%%%
%%%%%%%%%%%%%%%%%%%%%%%%%%%%%%%%%%%%%%%%%%%
\section{Discussion}
\label{section5}

The non-detection of diffuse radio emission in the CZ1359 field may be ascribed to some reasons. One of the potential explanations would be the collision impact parameter and viewing angle, both of which can alter apparent morphology (e.g., \cite{ay10}). While this geometrical dilution on radio halos and mini-halos are not well-known, relics can be significantly diluted if the shock surface is nearly perpendicular to the line-of-sight (LOS). However, \citet{kat15} discussed bulk velocities of member galaxies and concluded that the collision is undergoing almost parallel to the sky plane. Therefore, the geometrical effect could be minor.

At the X-ray shock front, the expansion speed of the post-shock region is estimated to be 2400~km/s \citep{kat15}. Since apparent thickness of the shock-heated region is $120''$ (168~kpc), the shock age is expected to be 68~Myr. This timescale is at least one order of magnitude shorter than the GeV electron cooling time \citep{sar99}. Therefore, the underluminous radio emission cannot be caused by synchrotron cooling.

A more realistic explanation would be that CRes are not accelerated and/or magnetic fields are not amplified enough, and hence synchrotron emission is too weak to be observed. In order to argue the possibilities, we discuss CRe content and magnetic-field strength below. For this purpose, we consider potential radio relic regions each with a size of $10'\times 3'$, as shown in Fig.~\ref{f05}(b). The red dashed box indicates the X-ray shock region \citep{kat15}, and the black solid boxes corresponds to potential late-stage shock regions. 

\begin{figure}[tp]
\begin{center}
\FigureFile(80mm,80mm){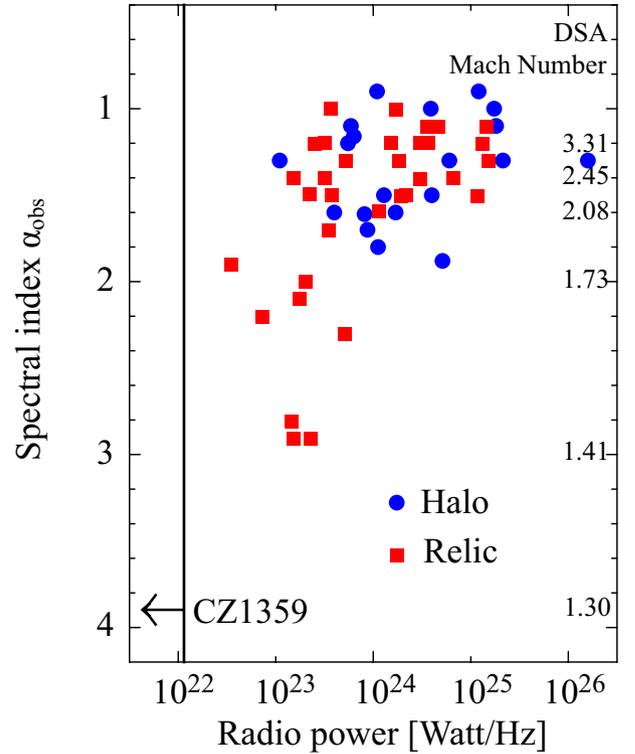}
\end{center}
\caption{
The correlation between the radio power at 1.4 GHz and the spectral index $\alpha_{\rm obs}$, all taken from \citet{fer12}. The black arrow indicates the constraint on CZ1359 from the present work (see text for details).
}
\label{f07}
\end{figure}

Suppose a simple power-law of the CRe energy spectrum and the standard formula of synchrotron radiation (Appendix B), we have the relations among the index of integrated (observed) radio spectra, $\alpha_{\rm obs}$, the index of injected radio spectra, $\alpha_{\rm inj}$, the shock Mach number, $M$ ($>1$), and the CRe energy spectral index, $p$, as follows:
\begin{equation}
\alpha_{\rm obs} = \alpha_{\rm inj} + \frac{1}{2} = \frac{M^2+1}{M^2-1} = \frac{p}{2},
\end{equation}
(e.g. \cite{be87}; \cite{tra15}). Fig.~\ref{f07} shows the $P_{\rm 1.4}$ -- $\alpha_{\rm obs}$ relation taken from \citet{fer12}. The vertical line indicates the upper limit of our observation, $P_{\rm 1.4}=1.14\times 10^{22}$ Watt/Hz. The arrow depicts $\alpha_{\rm obs} \sim 3.9$ derived from $M \sim 1.3$ of the X-ray shock, and this gives a very steep slope, $p=7.8$. Meanwhile, many observed relics have $\alpha_{\rm obs} \sim 1.0$ -- $1.5$. This corresponds to $M \sim 2$ -- 4 in the DSA picture. Such a range of the Mach number is supported by numerical simulations and is common for the late-stage shock. 

Under the assumption of the minimum energy equipartition (Appendix B), the steep spectrum of $p=7.8$ for the X-ray shock gives a very small CRe density and negligible intensity $I\sim 1.06\times 10^{-8}f^{-1}$~mJy/beam, where $f$ is the ratio of the cosmic-ray energy density over the cosmic-ray electron energy density. Therefore, the non-detection of diffuse radio emission at the X-ray shock front can be explained by the standard DSA scenario.

As a cross check, let us consider a case in which CZ1359 hosts a $\alpha \sim 1.5$ relic or halo, but is just dim. Our upper limit of 0.5 mJy/beam at 1.4 GHz and representative values ($p$, $\gamma_{\rm min}$, $\gamma_{\rm max}$, $D_{\rm los}$) = (3, 200, 3000, 1~Mpc) gives $B<0.682f$~${\rm \mu G}$, where $D_{\rm los}$ is the depth of the shock along the LOS. $\gamma_{\rm min}$ and $\gamma_{\rm max}$ are the minimum and maximum Lorentz factor of the cosmic-ray electrons, respectively. The upper limit is approximately proportional to $\gamma_{\rm min}^{-1}$ for $50\le \gamma_{\rm min} \le 600$, and changes by only a few \% for $\gamma_{\rm max}>3000$. The upper limit excludes the presence of strongly-magnetized shocks with $M \gtrsim 2$ in the standard DSA scenario. The upper limit also gives the limit of the cosmic-ray electron density, $n_{\rm ce} < 6.03\times 10^{-11}f$~${\rm cm^{-3}}$. With the mean thermal electron density at a potential late-stage shock, $n_{\rm e}\sim 0.5\times 10^{-3}$~${\rm cm^{-3}}$ \citep{kat15}, the injected fraction of CR electrons is $\xi_e = n_{\rm ce}/n_{\rm e} < 1.21f \times 10^{-7}$. This value is small compared to the efficiency estimated for a $M\sim 2$--4 shock, $\xi_e \sim 10^{-3}$--$10^{-6}$ (see e.g., \cite{kr13, vb14}). Finally, the upper limit gives rather loose upper limits of the non-thermal to thermal energy budget, $\varepsilon_{NT}/\varepsilon_{T} < 4.0f^2$~\% and $< 3.1f^2$~\% for the temperature $T_{\rm e}=4.6$~$\times 10^7$~K (North) and $T_{\rm e}=5.8$~$\times 10^7$~K (South), respectively \citep{kat15}. 

Note that we have used the Mach number measured from the X-ray observation. However, an inconsistency among the Mach numbers derived from X-ray and radio observations has been reported in the literature (e.g., \cite{ak13, ogr13, ita15}). Therefore, there might exist a discrepancy between radio-derived spectral index and X-ray-derived spectral index. 

Finally, let us consider the picture beyond the DSA. A simple DSA model assumes that cosmic-rays are well-accelerated with sufficient acceleration time. This may not be the case for the X-ray shock with an age of $68$~Myr \citep{kat15}, so that both field amplification and particle acceleration may be in progress, resulting that synchrotron emission is too weak to be detected. This scenario implies weak pre-amplified magnetic field and less seed cosmic-ray electrons at the linking region of the cluster. Nevertheless, low-energy cosmic-ray electrons, which emit radio synchrotron at low frequency and not in GHz band, may exist on some level. Thus, low frequency deep observation will be important as the next step. 

Recently, \citet{fuj15} and \citet{fuj16} proposed that cluster diffuse radio emission can be produced through the second-order Fermi (re-)acceleration, which is different from the DSA (the first-order Fermi acceleration, see also \cite{kan17}). Even if the second-order Fermi acceleration is taking place, electrons may not have had enough time to be accelerated, or scatterers of electrons (e.g. turbulence) may have not well developed behind the shock. Numerical simulations suggests that the early-stage merging clusters have relatively-weak turbulence, which is also consistent with non detection of radio halo in CZ1359.

%%%%%%%%%%%%%%%%%%%%%%%%%%%%%%%%%%%%%%%%%%%
%%%%%%%%%%%%%%%%%%%%%%%%%%%%%%%%%%%%%%%%%%%
\section{Conclusion}
\label{section6}

We conducted the ATCA 16 cm observation of a merging galaxy cluster, CIZA J1358.9-4750 (CZ1359). In the CZ1359 field, we first obtained a significant upper limit of diffuse emission, which was about one order of magnitude lower than the one estimated using the X-ray luminosity -- radio power relation for bright radio halos and relics. Therefore, an environment of this merging-cluster system is different from the clusters possessing typical, bright radio halos and relics. Using the upper limit of total intensity and assuming a model of shock acceleration, we first derived non-thermal properties at the X-ray shock front and potential shock fronts in CZ1359. 

Because a steep spectral index is expected due to a low ($\sim 1.3$) Mach number, low frequency deep observation with MWA and GMRT would be important to further investigate possible diffuse radio emission in the CZ1359 field. In future, the Square Kilometre Array (SKA) and its precursors, ASKAP and MeerKAT, will provide an unprecedented sensitivity in Southern hemisphere \citep{jo15}. They will advance the study of this cluster significantly. 

\vskip 12pt

The authors acknowledge Mr. Lijo Thomas George and Dr. Ruta Kale for their kind attentions and check of TGSS and GLEAM data. The authors would also like to thank the anonymous referee for his/her constructive comments and suggestions. This work was supported in part by JSPS KAKENHI Grant Numbers 26400218(MT), 15H03639(NK, TA), 15K05080(YF), 15K17614(TA), and by an University Research Support Grant from the National Astronomical Observatory of Japan (NAOJ). The Australia Telescope Compact Array is funded by the Commonwealth of Australia for operation as a National Facility managed by CSIRO.

\appendix 
%%%%%%%%%%%%%%%%%%%%%%%%%%%%%%%%%%%%%%%%%%%
%%%%%%%%%%%%%%%%%%%%%%%%%%%%%%%%%%%%%%%%%%%
\section*{A. Flux Scale}

\begin{figure}[tp]
\begin{center}
\FigureFile(70mm,70mm){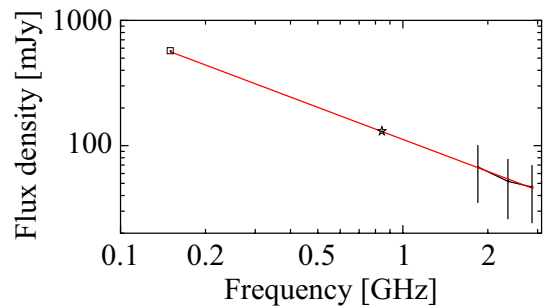}
\end{center}
\caption{
Total intensity spectrum of SUMSS J140033-473800. The black solid line with error bars show the ATCA 512 MHz data with CA06 and $r=2.0$. The square and star indicate the data of TGSS and SUMSS, respectively. The red line is the best-fit of the ATCA 512 MHz data.
}
\label{f08}
\end{figure}

Since there is no previous 16 cm observation of the CZ1359 field, we tried to verify our flux scale by comparing our result with the survey data of TGSS (150 MHz) and SUMSS (843 MHz). The comparison was performed for a bright compact source SUMSS J140033-473800 (14:00:33.5, -47:38:00) in the CZ1359 field. We carried out the linear fit for the spectrum using data at 1844, 2356, and 2868~MHz with each 512~MHz bandwidth. Here, the lowest band at 1332~MHz was not used for the fit, since it may contain large uncertainties derived from the data at $<1.5$~GHz (shown in Fig.~\ref{f01}). We used the images made with the robustness $r=2.0$, which was the one used to estimate the upper limit of diffuse emission in this paper. We find that the spectral index is $\alpha_{\rm obs} = 0.85$, which implies that the source is likely an unresolved Fanaroff-Riley radio galaxy rather than a quasar with a flat spectral index. If this is the case, the slope could extend to lower frequency bands (e.g., \cite{far14}). The extrapolation of the fit gives 129.76~mJy at 843~MHz and 562.88~mJy at 150 MHz, which are in agreement with 130.6~mJy at 843~MHz (SUMSS) and 573~mJy at 150 MHz (TGSS), respectively, on a few \% error level (Fig.~\ref{f08}). Therefore, we conclude that the flux scale of our observation is reasonable and our upper limit of diffuse emission is reliable. Note that we also performed such a comparison for images with a different robustness and a bandwidth, and obtained an uncertainty of typically $\sim 10$--20~\% on the flux scale.

\appendix 
%%%%%%%%%%%%%%%%%%%%%%%%%%%%%%%%%%%%%%%%%%%
%%%%%%%%%%%%%%%%%%%%%%%%%%%%%%%%%%%%%%%%%%%
\section*{B. Synchrotron Radiation from Cosmic-Ray Electrons}

We summarize the derivation of non-thermal properties referred in this paper. A similar derivation can be seen in, for example, \citet{gf04}. 

We consider that there are fresh CRes near a shock front and their energy spectrum follows a power law with an index $p$. We consider $p > 2$ to avoid an imaginary Mach number in the DSA picture, $M^2 = (p+2)/(p-2)$. The number density of CRes between the Lorentz factor $\gamma$ and $\gamma+d\gamma$ is defined as 
\begin{equation}
n_{\rm ce}(\gamma) d\gamma= N_0 \gamma^{-p} d\gamma,
\end{equation}
where $N_0$ is the normalization in units of ${\rm cm^{-3}}$. The number density of CRes is given by integrating the equation,
\begin{eqnarray}
n_{\rm ce}&=&N_0  \int_{\gamma_{\rm min}}^{\gamma_{\rm max}} \gamma^{-p} d\gamma \nonumber \\
&\equiv& N_0 N_1(p,\gamma_{\rm min},\gamma_{\rm max})~~~({\rm cm^{-3}})
\end{eqnarray}
where $N_1=[\gamma^{1-p}]_{\gamma_{\rm min}}^{\gamma_{\rm max}}/(1-p)$. Likewise, the energy density of CRes can be calculated as
\begin{eqnarray}
\varepsilon_{\rm ce} &=& N_0 m_{\rm e}c^2 \int_{\gamma_{\rm min}}^{\gamma_{\rm max}} \gamma^{1-p} d\gamma \nonumber \\
&\equiv& N_0 m_{\rm e}c^2 N_2(p,\gamma_{\rm min},\gamma_{\rm max})~~~({\rm erg/cm^3})
\end{eqnarray}
where $m_{\rm e}$ is the electron rest mass, $c$ is the speed of light, and  $N_2=[\gamma^{2-p}]_{\gamma_{\rm min}}^{\gamma_{\rm max}}/(2-p)$.

While $p$ is positive and steep, $n_{\rm ce}$ and $\varepsilon_{\rm ce}$ are insensitive to $\gamma_{\rm max}$ and we choose $\gamma_{\rm max} = 3000$. Actually, they change only a few \% as we vary $\gamma_{\rm max}$ from $3000$ to infinity in the case of $p = 3$. Meanwhile, they significantly depend on $\gamma_{\rm min}$. We choose $\gamma_{\rm min} = 200$ as a representative value, because CRes with $\gamma \sim 300$ have the longest cooling timescale and could be a dominant component in a typical cluster environment \citep{sar99}.

The CRe density can be constrained under the assumption that a radio emitter is in a simple minimum energy equipartition (see e.g., \cite{pfr04}). That is, the energy density of magnetic field, $\varepsilon_B$, is equal to the energy density of relativistic particles:
\begin{equation}\label{eq:equipartition}
f\varepsilon_{\rm ce} \equiv \varepsilon_{\rm ci} + \varepsilon_{\rm ce} = \varepsilon_B,
\end{equation}
where $\varepsilon_{\rm ci}$ is the energy density of cosmic-ray ions, and $f$ is the cosmic-ray energy fraction; $f=2$ $(\varepsilon_{\rm ci} : \varepsilon_{\rm ce} = 1 : 1)$ for example. With the assumption, we obtain
\begin{equation}
N_0 = \frac{B^2}{8\pi fm_{\rm e}c^2N_2}.
\end{equation}
\begin{equation}
n_{\rm ce} = \frac{N_1B^2}{8\pi fm_{\rm e}c^2N_2} \equiv N^*(p,\gamma_{\rm min},\gamma_{\rm max}) B^2,
\end{equation}
For example, $N_0\sim 1.0\times 10^{-5}$~${\rm cm^{-3}}$ and $N^*\sim 65$~${\rm cm^{-3}/G^2}$, if we consider ($f$, $p$, $\gamma_{\rm min}$, $\gamma_{\rm max}$) = (2, 3, 200, 3000).

The synchrotron emissivity depends on the strength of magnetic field perpendicular to the LOS, $B_{\perp}$. Specific Stokes parameters of synchrotron radiation can be written as (see \cite{rl79, wae09}), 
\begin{equation}
I = \int G_1(p) N_0(r) B_{\perp}(r)^{(1+p)/2} \omega^{(1-p)/2}dr,
\end{equation}
\begin{equation}
Q+iU=\int G_2(p) N_0(r) B_{\bot}(r)^{(1+p)/2} \omega^{(1-p)/2} e^{2i \chi (r)}dr,
\end{equation}
where $\chi$ is the initial polarization angle at $r$ and $\omega=2\pi \nu$ is the angular frequency. Other physical coefficients are $G_1(p)=2g_1(p)j(p)$, $G_2(p)=2g_2(p)j(p)$, 
\begin{equation}
j(p) = \frac{1}{4 \pi} \frac{\sqrt{3} e^3}{8 \pi m_{\rm e} c^2} \left(\frac{2m_{\rm e}c}{3e} \right)^{(1-p)/2}, 
\end{equation}
\begin{equation}
g_1(p) = \frac{1}{1+p} 2^{(1+p)/2} \Gamma \left(\frac{p}{4} - \frac{1}{12} \right)\Gamma \left(\frac{p}{4} + \frac{19}{12} \right),
\end{equation}
\begin{equation}
g_2(p) = 2^{(p-3)/2} \Gamma \left(\frac{p}{4} - \frac{1}{12} \right)\Gamma \left(\frac{p}{4} + \frac{7}{12} \right),
\end{equation}
where, $e$ is the electric charge and $\Gamma$ is the $\Gamma$ function. For a uniform medium with the LOS depth $D_{\rm los}$, equation~(A7) is integrated as 
\begin{equation}
I = \frac{G_1D_{\rm los}}{8\pi fm_{\rm e}c^2N_2}B_{\perp}^{(1+p)/2} B^2 \omega^{(1-p)/2}.
\end{equation}

We adopt a simple case where weak, pre-amplified intergalactic magnetic field is just compressed by the shock, hence the strength of magnetic field perpendicular to the LOS is the same as that of magnetic field parallel to, and the component parallel to the shock normal is negligibly small. It means $B=\sqrt{2} B_{\perp}$. The intensity is written as
\begin{eqnarray}
I &=&  I^*(p,\gamma_{\rm min},\gamma_{\rm max}) B^{(5+p)/2} \omega^{(1-p)/2} \nonumber \\
&=& \frac{2^{-(1+p)/4}G_1D_{\rm los}}{8\pi fm_{\rm e}c^2N_2}B^{(5+p)/2} \omega^{(1-p)/2}.
\end{eqnarray}

The thermal energy density, $\varepsilon_T$, and $\varepsilon_B$ are respectively given in ${\rm erg/cm^{3}}$ by
\begin{equation}
\varepsilon_T = \frac{3}{2}nkT = 4.00 \times 10^{-12} \left(\frac{n_{\rm e}}{10^{-3}~{\rm cm^{-3}}}\right) \left(\frac{T_{\rm e}}{10^7~{\rm K}}\right),
\end{equation}
\begin{equation}
\varepsilon_B = \frac{B^2}{8\pi} = 3.98 \times 10^{-14} \left(\frac{B}{\mu{\rm G}}\right)^2,
\end{equation}
where $n$ and $T$ are the ICM density and temperature, respectively. The ICM is assumed to be in thermal equilibrium ($T=T_{\rm e}$, $T_{\rm e}$ is the electron temperature) and fully-ionized with the mass fractions of Hydrogen $X=0.76$ and of Helium $Y=0.24$. We can then estimate the energy budget of the nonthermal components under the energy equipartition,
\begin{equation}
\varepsilon_{NT} = \varepsilon_{\rm ci} + \varepsilon_{\rm ce} + \varepsilon_B = 2 \varepsilon_B
\end{equation}
\begin{equation}
\frac{\varepsilon_{NT}}{\varepsilon_{T}} < 1.99~\% \left(\frac{B}{\rm 1~\mu G} \right)^{2}\left(\frac{10^{-3}~{\rm cm^{-3}}}{n_{\rm e}}\right)
\left(\frac{10^7~{\rm K}}{T_{\rm e}}\right).
\end{equation}


\begin{thebibliography}{}
\bibitem[Akahori \& Yoshikawa(2008)]{ay08}
	Akahori, T., Yoshikawa, K. 2008, PASJ, 60, L19
\bibitem[Akahori \& Yoshikawa(2010)]{ay10}
	Akahori, T., Yoshikawa, K. 2010, PASJ, 62, 335
\bibitem[Akahori \& Yoshikawa(2012)]{ay12}
	Akahori, T., Yoshikawa, K. 2012, PASJ, 64, 12
\bibitem[Akamatsu \& Kawahara(2013)]{ak13}
	Akamatsu, H., \& Kawahara, H. 2013, PASJ, 65, 16
\bibitem[Akamatsu et al.(2016)]{ak16}
	Akamatsu, H., Gu, L., Shimwell, T. W., Mernier, F., Mao, J., Urdampilleta, I., de Plaa, J.,\& R\"{o}ttgering, H. J. A. 2016, A\&A, 593, L7
\bibitem[Blandford \& Eichler(1987)]{be87}
	Blandford, R., \& Eichler, D. 1987, Phys. Rep., 154, 1
\bibitem[Bock et al.(1999)]{bock99}
	Bock, D., Large, M. I., \& Sadler, E. M. 1999, AJ, 117, 1578
\bibitem[Bonafede et al.(2014)]{bon14}
	Bonafede, A., Intema, H. T., Br\"{u}ggen, M., et al. 2014, MNRAS, 444, 44
\bibitem[Brunetti et al.(2007)]{bru07}
	Brunetti, G., Venturi, T., Dallacasa., D., et al. 2007, ApJL, 670, L5
\bibitem[Brunetti et al.(2009)]{bru09}
	Brunetti, G., Cassano, R., Dolag, K., \& Setti, G., 2009, A\&A, 507, 661
\bibitem[Caprioli \& Spitkovsky(2014)]{cs14}
	Caprioli, D., \& Spitkovsky, A. 2014, ApJ, 783, 91
\bibitem[Carilli \& Taylor(2002)]{ct02}
	Carilli, C. L., \& Taylor, G. B. 2002, ARA\&A, 40, 319
\bibitem[Cassano et al.(2010)]{cas10}
	Cassano, R., Ettori, S., Giacintucci, S., et al. 2010, ApJL, 721, L82
\bibitem[Cassano et al.(2013)]{cas13}
	Cassano, R., Ettori, S., Brunetti, G., Giacintucci, S., Pratt, G. W., Venturi, T., Kale, R., Dolag, K., \& Markevitch, M. 2013, ApJ, 777, 141
\bibitem[Dawson(2013)]{daw13}
	Dawson, W. A. 2013, ApJ, 772, 131
\bibitem[Dawson(2015)]{daw15}
	Dawson, W. A., et al. 2015, ApJ, 805, 143
\bibitem[de Gasperin et al.(2014)]{deg14}
	de Gasperin, F., van Weeren, R. J., Br\"{u}ggen, M., Vazza, F., Bonafede, A., \& Intema, H. T. 2014, MNRAS, 444, 3130
\bibitem[Ebeling et al.(2002)]{ebe02}
	Ebeling, H., Mullis, C. R, \& Tully, R. B. 2002, ApJ, 580, 774
\bibitem[En{\ss}lin et al.(1998)]{en98}
	En{\ss}lin, T. A., Biermann, P. L., Klein, U., \& Kohle, S. 1998, A\&A, 332, 395
\bibitem[En{\ss}lin et al.(2011)]{ens11}
	En{\ss}lin, T., Pfrommer, C., Miniati, F., \& Subramanian, K. 2011, A\&A, 527, A99
\bibitem[Farnes et al.(2014)]{far14}
	Farnes, J. S., Gaensler, B. M., \& Carretti, E. 2014, ApJS, 212, 15
\bibitem[Feretti et al.(2012)]{fer12}
	Feretti, L., et al. 2012, A\&A Review, 20, 54
\bibitem[Finoguenov et al.(2010)]{fi10}
	Finoguenov, A., Sarazin, C. L., Nakazawa, K., Wik, D. R., \& Clarke, T. E. 2010, ApJ, 715, 1143
\bibitem[Fujita et al.(2008)]{fuj08}
	Fujita, Y., et al. 2008, PASJ, 60, S343
\bibitem[Fujita et al.(2015)]{fuj15}
	Fujita, Y., Takizawa, M., Yamazaki, R., Akamatsu, H., \& Ohno, H. 2015, ApJ, 815, 116
\bibitem[Fujita et al.(2016)]{fuj16}
	Fujita, Y., Akamatsu, H., \& Kimura, S. S. 2016, PASJ, 68, 34
\bibitem[Giovannini et al.(2011)]{gio11}
	Giovannini, G., Feretti, L., Girardi, M., Govoni, F., Murgia, M., Vacca, V.,\& Bagchi, J. 2011, A\&A, 530, 5
\bibitem[Gitti et al.(2015)]{git15}
	Gitti, M., Tozzi, P., Brunetti, G., et al. 2015, aska.conf, 76
\bibitem[Govoni \& Feretti(2004)]{gf04}
	Govoni, F., \& Feretti, L. 2004, International Journal of Modern Physics D, 13, 1549
\bibitem[Govoni et al.(2013)]{gov13}
	Govoni, F., Murgia, M., Xu, H., Li, H., Norman, M. L., Feretti, L., Giovannini, G., \& Vacca, V. 2013, A\&A, 554, 102
\bibitem[Guo et al.(2014)]{guo14}
	Guo, Xinyi, Sironi, L., \& Narayan, R., 2014, ApJ, 794, 153
\bibitem[Hurley-Walker et al.(2017)]{hur17}
	Hurley-Walker, N., Callingham, J. R., Hancock, P. J., et al., 2017, MNRAS, 464, 1146
\bibitem[Intena et al.(2017)]{int17}
	Intema, H. T., Jagannathan, P., Mooley, K. P., \& Frail, D. A. 2017, A\&A, 598, 78
\bibitem[Itahana et al.(2015)]{ita15}
	Itahana, M., Takizawa, M., Akamatsu, H., Ohashi, T., Ishisaki, Y., Kawahara, H., \& Van Weeren, R. J. 2015, PASJ, 67, 113
\bibitem[Jee et al.(2016)]{jee16}
	Jee, M., Dawson, W. A., Stroe, A., Wittman, D., van Weeren, R. J., Br\"{u}ggen, M., Bradac, M., R\"{o}ttgering, H. 2016, ApJ, 817, 179
\bibitem[Johnston-Hollitt et al.(2015)]{jo15}
	Johnston-Hollitt, M., Govoni, F., Beck, R., et al. 2015, in Proc. of Advancing Astrophysics with the Square Kilometre Array (Trieste: SISSA), PoS (AASKA14)092, id.92
\bibitem[Kang \& Ryu(2013)]{kr13}
	Kang, H., \& Ryu, D. 2013, ApJ, 764:95
\bibitem[Kang et al.(2017)]{kan17}
	Kang, H., Ryu, D., \& Jones, T. W. 2017, ApJ, 840, 42
\bibitem[Kale et al.(2015)]{kal15}
	Kale, R., Venturi, T., Giacintucci, S., Dallacasa, D., Cassano, R., Brunetti, G., Cuciti, V., Macario, G., \& Athreya, R. 2015, A\&A, 579, A92
\bibitem[Kato et al.(2015)]{kat15}
	Kato, Y., Nakazawa, K., et al. 2015, PASJ, 67, 71
%\bibitem[Kazemi et al.(2011)]{kaz11}
%	Kazemi, S., Yatawatta, S., Zaroubi, S., et al.\ 2011, \mnras, 414, 1656 
\bibitem[Kitayama et al.(2016)]{kit16}
	Kitayama, T., Ueda, S., Takakuwa, T., et al. (2016), PASJ, 68, 88
\bibitem[Kocevski et al.(2007)]{koc07}
	Kocevski, D. D., Ebeling, H., Mullis, C. R., \& Tully, R. B. 2007, ApJ, 662, 224
\bibitem[Markevitch, Viklinin(2007)]{mar07}
	Markevitch, M., \& Vikhlinin, A. 2007, Physics Reports, 443, 1
\bibitem[Matsukiyo et al.(2011)]{ma11}
	Matsukiyo, S., Ohira, Y., Yamazaki, R., \& Umeda, T.\ 2011, ApJ, 742, 47
\bibitem[Matsukiyo \& Matsumoto(2015)]{ma15}
	Matsukiyo, S. \& Matsumoto, Y.\ 2015, JPhCS, 642a, 2017
\bibitem[Nakazawa et al.(2009)]{nak09}
	Nakazawa, K. et al. 2009, PASJ, 61, 339
\bibitem[Ogrean et al.(2013)]{ogr13}
	Ogrean, G. A., Br\"{u}ggen, M., van Weeren, R. J., R\"{o}ttgering, H., Croston, J. H., \& Hoeft, M. 2013, MNRAS, 433, 812
\bibitem[Okabe \& Umetsu(2008)]{oka08}
	Okabe, N., \& Umetsu, K. 2008, PASJ, 60, 345
\bibitem[Okabe et al.(2011)]{oka11}
	Okabe, N., Bourdin, H., Mazzotta, P., \& Maurogordato, S. 2011, ApJ, 741, 116
\bibitem[Okabe et al.(2015)]{oka15}
	Okabe, N., Akamatsu, H., Kakuwa, J., Fujita, Y., Zhang, Y., Tanaka, M., \& Umetsu, K. 2015, PASJ, 67, 114
\bibitem[Ozawa et al.(2015)]{oza15}
	Ozawa, T.,  Nakanishi, H., Akahori, T., Anraku, K., Takizawa, M., Takahashi, I., Onodera, S., Tsuda, Y., \& Sofue, Y. 2015, PASJ, 67, 110
\bibitem[Pfrommer \& En{\ss}lin(2004)]{pfr04}
	Prfommer, C., \& En{\ss}lin, T. A. 2004, MNRAS, 352, 76
\bibitem[Rudnick, Lemmerman(2009)]{rud09}
	Rudnick, L., \& Lemmerman, J. 2009, ApJ, 697, 1341
\bibitem[Rybicki \& Lightman(1979)]{rl79}
	Rybicki, G. B., \& Lightman, A. P. 1979, Radiation processes in astrophysics (New York: Wiley-Interscience)
\bibitem[Ryu et al.(2003)]{ryu03}
	Ryu, D., Kang, H., Hallman, E., \& Jones, T. W. 2003, ApJ, 593, 599
\bibitem[Sarazin(1986)]{sar86}
	Sarazin, C. L. 1986, Reviews of Modern Physics, 58, 1
\bibitem[Sarazin(1999)]{sar99}
	Sarazin, C. L. 1999, ApJ, 520, 529
\bibitem[Sarazin(2002)]{sar02}
	Sarazin, C. L. (2002), ASSL, 272, 1
\bibitem[Takizawa(2008)]{tak08}
	Takizawa, M. 2008, ApJ, 687, 951
\bibitem[Trasatti et al.(2015)]{tra15}
	Trasatti, M., Akamatsu, H., Lovisari, L., Klein, U., Bonafede, A., Br{\"u}ggen, M., Dallacasa, D., \& Clarke, Tracy. 2015, A\&A, 575, 45
\bibitem[Vazza \& Br\"uggen(2014)]{vb14}
	Vazza, F., \& Br\"uggen, M. 2014, MNRAS, 437, 2291
\bibitem[van Weeren et al.(2017)]{van17}
	van Weeren, R. J., et al. 2017, Nature Astronomy, 1, 5
\bibitem[Waelkens et al.(2009)]{wae09}
	Waelkens, A., Jaffe, T., Reinecke, M., Kitaura, F. S., \& En{\ss}lin, T. A., 2009, \aap, 495, 697
\bibitem[Wilson et al.(2011)]{wil11}
	Wilson, E., et al. 2011, MNRAS, 416, 832
\bibitem[Xu et al.(2012)]{xu12}
	Xu, H., Govoni, F., Murgia, M., Li, H., Collins, D. C., Norman, M. L., Cen, R., Feretti, L., \& Giovannini, G. 2012, ApJ, 759, 40
\end{thebibliography}
\end{document}